\documentclass[reprint,
 amsmath,amssymb,
 aps,
]{revtex4-2}

\usepackage{graphicx}
\usepackage{dcolumn}
\usepackage{bm}
\usepackage{xcolor}

\begin{document}


\title{Realizing normal group-velocity dispersion in free space via angular dispersion}

\author{Layton A. Hall$^{1}$}
\author{Ayman F. Abouraddy$^{1,*}$}
\affiliation{$^{1}$CREOL, The College of Optics \& Photonics, University of Central~Florida, Orlando, FL 32816, USA}
\affiliation{$^*$Corresponding author: raddy@creol.ucf.edu}

\begin{abstract}
It has long been thought that normal group-velocity dispersion (GVD) cannot be produced in free space via angular dispersion. Indeed, conventional diffractive or dispersive components such as gratings or prisms produce only anomalous GVD. We identify the conditions that must be fulfilled by the angular dispersion introduced into a plane-wave pulse to yield normal GVD. We then utilize a pulsed-beam shaper capable of introducing arbitrary angular-dispersion profiles to symmetrically produce normal and anomalous GVD in free space, which are realized here on the same footing for the first time.
\end{abstract}


\maketitle

Angular dispersion (AD) refers to the wavelength-dependent propagation angle inculcated into polychromatic fields by diffractive or dispersive components such as gratings or prisms \cite{Torres10AOP,Fulop10Review}. Because AD can change the group velocity of a pulse and introduce group-velocity dispersion (GVD), it has found myriad applications in dispersion compensation \cite{Fork84OL,Gordon84OL}, pulse compression \cite{Lemoff93OL,Kane97JOSAB}, and facilitating nonlinear interactions \cite{Hebling02OE}. In the visible and near-infrared, anomalous GVD produced by AD can counterbalance normal GVD in optical materials \cite{Szatmari90AO,Szatmari96OL,Sonajalg96OL,Sonajalg97OL}. However, AD does not produce anomalous and normal GVD symmetrically. In fact, the normal GVD required to compensate for anomalous GVD at longer wavelengths beyond the zero-dispersion wavelength has not been realized to date. Indeed, a well-known result by Martinez, Gordon, and Fork (henceforth `MGF') \cite{Martinez84JOSAA} purports to prove that AD can produce \textit{only} anomalous GVD in free space.

A counterexample to the MGF result was proposed by Porras \textit{et al}. \cite{Porras03PRE2} in which a particular AD profile produces normal GVD, but it has not been realized experimentally to date. The existence of propagation-invariant wave packets in the anomalous GVD regime \cite{Zamboni03OC,Longhi03PRE,Longhi04OL,Porras04PRE,Malaguti08OL} has been established theoretically, which implies that they incorporate AD-driven normal GVD, but no proposals have been offered for their realization. From a fundamental perspective, it is not clear why such counterexamples exist in contradistinction to MGF. In other words, what are the implicit assumptions that delimit the domain of validity of MGF? And how can AD-driven normal GVD be realized in free space?

We show here that the MGF result is \textit{not} universally valid, and AD can indeed readily produce normal GVD in free space. Synthesizing such field configurations, however, requires either independent control over multiple orders of AD, or the ability to introduce \textit{non-differentiable} AD \cite{Hall21OL,Yessenov21ACSP}. That is, a wavelength-dependent propagation angle that does \textit{not} possess a derivative at some wavelength, a phenomenon that we recently showed undergirds the unique characteristics of `space-time' (ST) wave packets \cite{Kondakci16OE,Parker16OE,Kondakci17NP,Porras17OL,Wong17ACSP2,Efremidis17OL,Yessenov19OPN,Yessenov19OE,Yessenov19PRA}. Conventional components such as gratings or prisms do not produce these forms of AD required to yield normal GVD. Instead, we utilize here a pulsed-beam shaper capable of introducing arbitrary AD profiles (both differentiable \textit{and} non-differentiable) into a plane-wave pulse, and observe for the first time normal and anomalous GVD in free space produced by AD on the same footing.

\begin{figure}[t!]
\centering
\includegraphics[width=8.6cm]{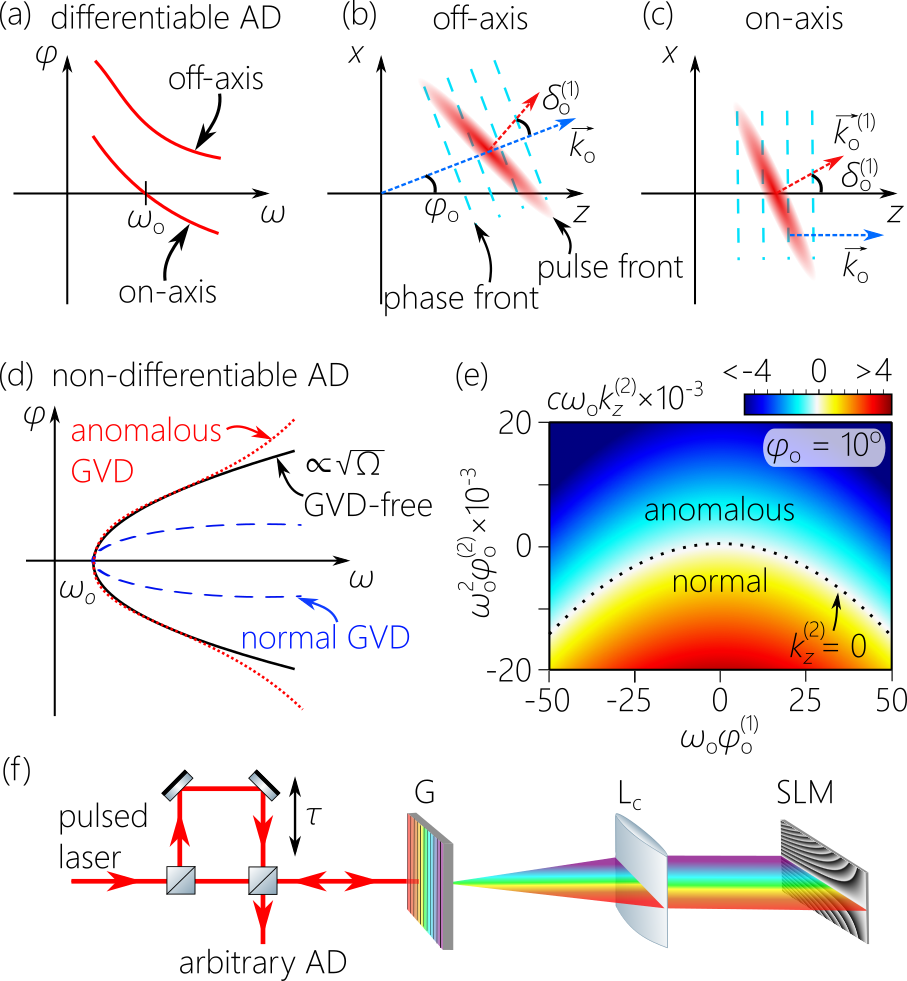}
\caption{(a) The propagation angle $\varphi(\omega)$ for on-axis and off-axis fields endowed with differentiable AD. (b) Off-axis and (c) on-axis field configurations. (d) Non-differentiable AD enabling GVD-free propagation, normal, or anomalous GVD for on-axis fields. (e) Calculated GVD coefficient $k_{z}^{(2)}$ from Eq.~\ref{Eq:AxialGVDCoefficient} as a function $\omega_{\mathrm{o}}\varphi_{\mathrm{o}}^{(1)}$ and $\omega_{\mathrm{o}}^{2}\varphi_{\mathrm{o}}^{(2)}$ for an off-axis field $\varphi_{\mathrm{o}}\!=\!10^{\circ}$. (f) Setup for realizing arbitrary AD profiles. G: Diffraction grating; L$_\mathrm{c}$: cylindrical lens; SLM: spatial light modulator.}
\label{Fig:ConceptSetup}
\end{figure}

We start with a scalar plane-wave pulse $E(x,z;t)\!=\!E(z;t)\!=\!\int\!d\omega\widetilde{E}(\omega)e^{-i\omega(t-z/c)}\!=\!E(0;t-z/c)$ traveling in free space at a group velocity $\widetilde{v}\!=\!c$; here $\widetilde{E}(\omega)$ is the Fourier transform of $E(0;t)$, $x$ and $z$ are the transverse and axial coordinates, respectively, and $c$ is the speed of light in vacuum. Introducing AD in the form of a frequency-dependent propagation angle $\varphi(\omega)$ [Fig.~\ref{Fig:ConceptSetup}(a)] yields $E(x,z;t)\!=\!\int\!d\omega\widetilde{E}(\omega)e^{ik(z\cos{\varphi}+x\sin{\varphi}-ct)}$ [Fig.~\ref{Fig:ConceptSetup}(b,c)], with transverse and axial wave numbers $k_{x}\!=\!k\sin\{\varphi(\omega)\}$ and $k_{z}\!=\!k\cos\{\varphi(\omega)\}$, respectively, where $k\!=\!\tfrac{\omega}{c}$. We expand $\varphi$ and $k_{z}$ into Taylor series around a carrier frequency $\omega_{\mathrm{o}}$ as $\varphi(\omega)\!=\!\varphi(\omega_{\mathrm{o}}+\Omega)\!=\!\varphi_{\mathrm{o}}+\varphi_{\mathrm{o}}^{(1)}\Omega+\tfrac{1}{2}\varphi_{\mathrm{o}}^{(2)}\Omega^{2}+\cdots$, and $k_{z}(\omega)\!=\!k_{z}^{(0)}+k_{z}^{(1)}\Omega+\tfrac{1}{2}k_{z}^{(2)}\Omega^{2}+\cdots$, where $\varphi_{\mathrm{o}}\!=\!\varphi(\omega_{\mathrm{o}})$, $\varphi_{\mathrm{o}}^{(n)}\!=\!\tfrac{d^{n}\varphi}{d\omega^{n}}\big|_{\omega_{\mathrm{o}}}$, $k_{z}^{(0)}\!=\!k_{\mathrm{o}}\cos{\varphi_{\mathrm{o}}}$, $k_{\mathrm{o}}\!=\!\omega_{\mathrm{o}}/c$, and the higher-order terms are:
\begin{eqnarray}
ck_{z}^{(1)}\!\!\!\!\!&=&\!\!\!\!\!\cos{\varphi_{\mathrm{o}}}-\omega_{\mathrm{o}}\varphi_{\mathrm{o}}^{(1)}\sin{\varphi_{\mathrm{o}}},\label{Eq:AxialFirstOrderWaveNumber}\\
c\omega_{\mathrm{o}}k_{z}^{(2)}\!\!\!\!\!&=&\!\!\!\!\!-(\omega_{\mathrm{o}}\varphi_{\mathrm{o}}^{(1)})^{2}\cos{\varphi_{\mathrm{o}}}\!-\!(\omega_{\mathrm{o}}^{2}\varphi_{\mathrm{o}}^{(2)}\!+\!2\omega_{\mathrm{o}}\varphi_{\mathrm{o}}^{(1)})\sin{\varphi_{\mathrm{o}}}\label{Eq:AxialGVDCoefficient}.
\end{eqnarray}
A similar series can be obtained for the transverse wave number $k_{x}(\omega)\!=\!k_{x}^{(0)}+k_{x}^{(1)}\Omega+\tfrac{1}{2}k_{x}^{(2)}\Omega^{2}+\cdots$. The phase front (the plane of constant phase) is orthogonal to the vector $\vec{k}_{\mathrm{o}}\!=\!(k_{x}^{(0)},k_{z}^{(0)})$, which makes an angle $\varphi_{\mathrm{o}}$ with the $z$-axis; the pulse front (the plane of constant amplitude) is orthogonal to the vector $\vec{k}_{\mathrm{o}}^{(1)}\!=\!(k_{x}^{(1)},k_{z}^{(1)})$, which makes an angle $\delta_{\mathrm{o}}^{(1)}$ with $\vec{k}_{\mathrm{o}}$, where $\tan{\delta_{\mathrm{o}}^{(1)}}\!=\!\omega_{\mathrm{o}}\varphi_{\mathrm{o}}^{(1)}$ [Fig.~\ref{Fig:ConceptSetup}(b,c)] \cite{Hebling96OQE,Porras03PRE2}. Throughout, the $z$-axis is the observation axis, along which the phase velocity is $v_{\mathrm{ph}}\!=\!c/\cos{\varphi_{\mathrm{o}}}$ and the group velocity is $\widetilde{v}\!=\!c/\{\cos{\varphi_{\mathrm{o}}}-\omega_{\mathrm{o}}\varphi_{\mathrm{o}}^{(1)}\sin{\varphi_{\mathrm{o}}}\}\!=\!c\cos{\delta^{(1)}}/\cos{(\delta^{(1)}+\varphi_{\mathrm{o}})}$, and the GVD coefficient is $k_{z}^{(2)}$ given in Eq.~\ref{Eq:AxialGVDCoefficient}. We distinguish between two field configurations: (1) on-axis fields where all the wavelengths propagate close to the $z$-axis, $\varphi_{\mathrm{o}}\!=\!0$ [Fig.~\ref{Fig:ConceptSetup}(c)], which are more appropriate for applications that require a significant propagation distance; and (2) off-axis fields where all the wavelengths travel at an angle with the $z$-axis, $\varphi_{\mathrm{o}}\!\neq\!0$ [Fig.~\ref{Fig:ConceptSetup}(b)], which are useful for interactions with localized structures.

We are now in a position to formulate the MGF result \cite{Martinez84JOSAA}. Starting with the axial GVD $k_{z}^{(2)}$ for an off-axis field (Eq.~\ref{Eq:AxialGVDCoefficient}), MGF align $\vec{k}_{\mathrm{o}}$ with the $z$-axis, whereupon $c\omega_{\mathrm{o}}k_{z}^{(2)}\!=\!-(\omega_{\mathrm{o}}\varphi_{\mathrm{o}}^{(1)})^{2}$, so the GVD is anomalous independently of the sign of $\varphi_{\mathrm{o}}^{(1)}$, and $\widetilde{v}\!=\!c$. Hence, this result applies only to on-axis fields. We show below that normal GVD can be produced via AD in off-axis fields, and even in on-axis fields by introducing non-differentiable AD. As such, the MGF result is valid only for on-axis fields endowed with differentiable AD.

We first examine GVD for on-axis fields. The results given above take for granted that $\varphi(\omega)$ is differentiable at $\omega\!=\!\omega_{\mathrm{o}}$, so that $\varphi_{\mathrm{o}}^{(1)}$ and $\varphi_{\mathrm{o}}^{(2)}$ are well-defined. Consider instead AD of the form $\varphi(\omega)\!\approx\!\eta\sqrt{\tfrac{\Omega}{\omega_{\mathrm{o}}}}$ for small $\varphi$ and constant $\eta$, which is \textit{not} differentiable at $\omega\!=\!\omega_{\mathrm{o}}$ ($\Omega\!=\!0$) [Fig.~\ref{Fig:ConceptSetup}(d)]. In this case $\widetilde{v}\!=\!c/\widetilde{n}$, $k_{z}^{(2)}\!=\!0$, and the group index is $\widetilde{n}\!=\!1-\tfrac{1}{2}\eta^{2}$. To produce GVD, it can be readily shown that modifying the AD to take the form:
\begin{eqnarray}\label{Eq:ADForNonDifferentiableAD}
&\sin&\!\!\!\!\!\!{\{\varphi(\omega)\}}=\eta\sqrt{\frac{\Omega}{\omega_{\mathrm{o}}}}\;\frac{\omega_{\mathrm{o}}}{\omega}\nonumber\\&\times&\!\!\!\!\!\sqrt{\left\{1+\frac{1+\widetilde{n}}{2}\frac{\Omega}{\omega_{\mathrm{o}}}+\frac{\sigma}{2}\left(\frac{\Omega}{\omega_{\mathrm{o}}}\right)^{2}\right\}\left\{1-\frac{\sigma}{1-\widetilde{n}}\frac{\Omega}{\omega_{\mathrm{o}}}\right\}},
\end{eqnarray}
results in $k_{z}\!=\!k_{\mathrm{o}}+\tfrac{\widetilde{n}}{c}\Omega+\tfrac{1}{2}k_{2}\Omega^{2}$, where $\sigma\!=\!\tfrac{1}{2}k_{2}\omega_{\mathrm{o}}c$, $\widetilde{v}\!=\!c/\widetilde{n}$, the GVD coefficient is $k_{z}^{(2)}\!=\!k_{2}$, and all the higher-order terms vanish. Crucially, this formulation is agnostic with respect to the sign of $k_{2}$: normal $k_{2}\!>\!0$ and anomalous $k_{2}\!<\!0$ are treated on the same footing. However, because of the $\sqrt{\Omega}$ term in Eq.~\ref{Eq:ADForNonDifferentiableAD}, realizing normal GVD on-axis in free space requires non-differentiable AD and $\widetilde{v}\!\neq\!c$.

\begin{figure*}[t!]
\centering
\includegraphics[width=17.6cm]{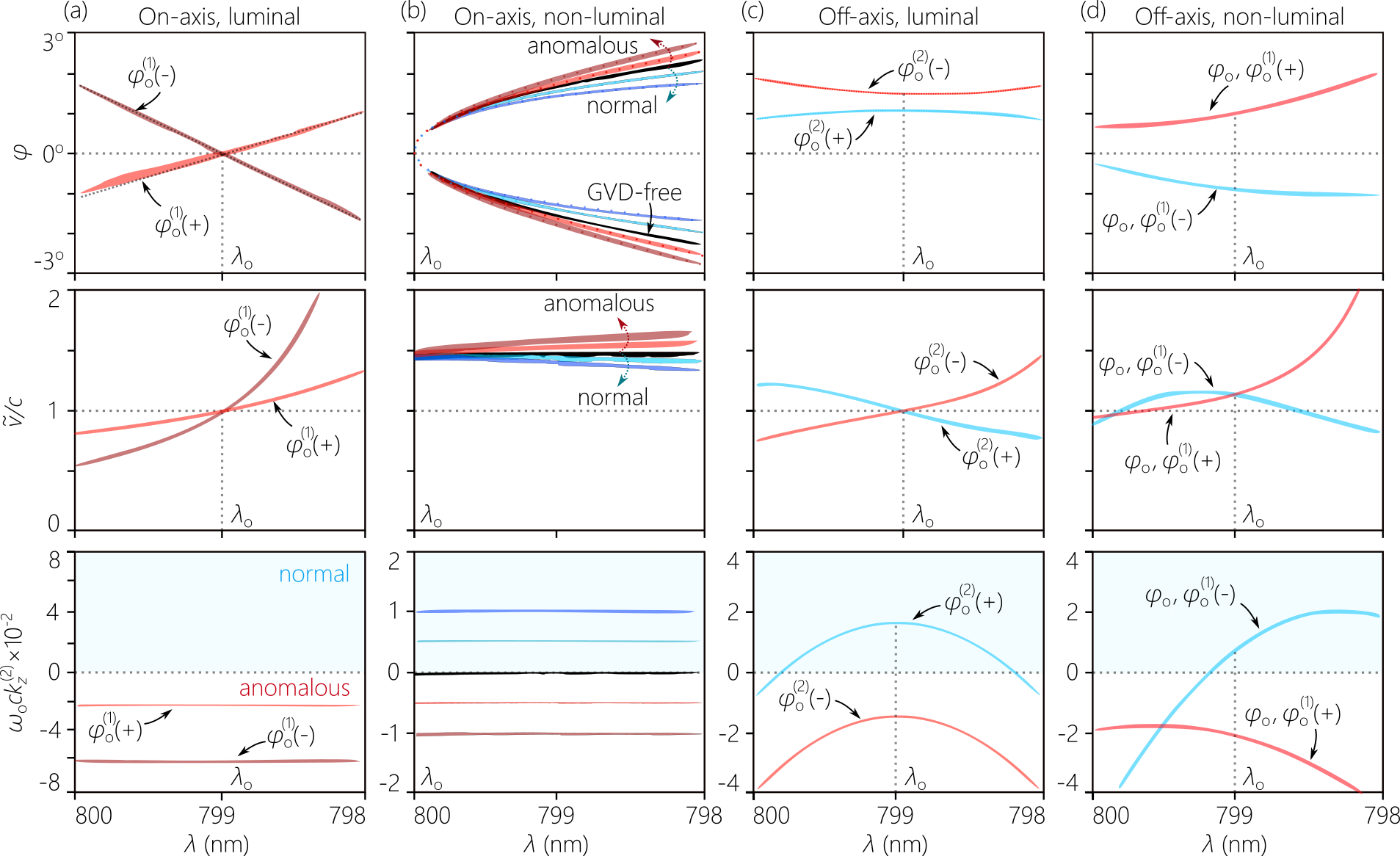}
\caption{Measurements of the propagation angle $\varphi(\lambda)$ in the first row, from which we extract the group velocity $\widetilde{v}(\lambda)\!=\!(k_{z}(\lambda)-k_{\mathrm{o}})/\Omega$ in the second row, and the GVD coefficient $k_{z}^{(2)}(\lambda)\!=\!2(k_{z}(\lambda)-k_{\mathrm{o}}-\Omega/\widetilde{v})/\Omega^2$ in the third row. The red curves correspond to anomalous GVD and the blue to normal GVD. (a) On-axis ($\varphi_{\mathrm{o}}\!=\!0$) luminal ($\widetilde{v}\!=\!c$) fields, with $\lambda_{\mathrm{o}}\!=\!799$~nm, and $\omega_{\mathrm{o}}\varphi^{(1)}_{\mathrm{o}}\!=\!10$ and $-25$. The GVD is anomalous in both cases. (b) Non-differentiable AD from Eq.~\ref{Eq:ADForNonDifferentiableAD} producing normal and anomalous GVD in on-axis fields with $\widetilde{v}\!=\!1.5c$, and $\omega_{\mathrm{o}}ck_{z}^{(2)}\!=\!\pm50$, $\pm100$, and 0 (GVD-free, black curve in the center). Propagation angles $\varphi\!\approx\!0^{\circ}$ are blocked by a spatial filter that removes the zero diffraction order, and the dotted curve is the theoretical prediction. (c) Off-axis luminal fields with $(\varphi_{\mathrm{o}},\omega_\mathrm{o}\varphi^{(1)}_{\mathrm{o}},\omega_\mathrm{o}^2\varphi^{(2)}_{\mathrm{o}})\!=\!(1^{\circ},-8.72\!\times\!10^{-3},-10^{4})$ in blue for normal GVD and $(1.5^{\circ},-1.3\!\times\!10^{-2},10^{4})$ in red for anomalous GVD. (d) Off-axis non-luminal fields with $(-1^{\circ},-8.72,10^{4})$ in blue for normal GVD and $(1^{\circ},8.72,10^{4})$ in red for anomalous GVD.}
\label{Fig:MeasuredGVD}
\end{figure*}

For off-axis fields, on the other hand, differentiable AD suffices to achieve normal GVD. Porras \textit{et al}. in \cite{Porras03PRE2} propose setting $\varphi_{\mathrm{o}}^{(1)}\!=\!0$ in Eq.~\ref{Eq:AxialFirstOrderWaveNumber} and Eq.~\ref{Eq:AxialGVDCoefficient}, whereupon $\widetilde{v}\!=\!c/\cos{\varphi_{\mathrm{o}}}$ and $c\omega_{\mathrm{o}}k_{z}^{(2)}\!=\!-\omega_{\mathrm{o}}^{2}\varphi_{\mathrm{o}}^{(2)}\sin{\varphi_{\mathrm{o}}}$, whose sign is determined by $\varphi_{\mathrm{o}}$ and $\varphi_{\mathrm{o}}^{(2)}$. More generally, normal GVD can be realized by judicious choice of $\varphi_{\mathrm{o}}$, $\varphi_{\mathrm{o}}^{(1)}$, and $\varphi_{\mathrm{o}}^{(2)}$, as is clear from Fig.~\ref{Fig:ConceptSetup}(e). Although there always exist normal- \textit{and} anomalous-GVD regions, the normal-GVD domain shrinks as $\varphi_{\mathrm{o}}\!\rightarrow\!0$. Off-axis fields exhibiting normal GVD may have $\widetilde{v}\!\neq\!c$ (as in \cite{Porras03PRE2}) or $\widetilde{v}\!=\!c$, the latter of which requires satisfying the constraint $\varphi_{\mathrm{o}}\!=\!-2\delta_{\mathrm{o}}^{(1)}$. In general, producing normal GVD in an off-axis field requires independent control over the first three AD terms $\varphi_{\mathrm{o}}$, $\varphi_{\mathrm{o}}^{(1)}$, and $\varphi_{\mathrm{o}}^{(2)}$. Conventional optical components such as gratings and prisms do not offer such tailored control; indeed, gratings produce solely anomalous GVD. Finally, the higher-order dispersion coefficients $k_{z}^{(n)}$ for $n\!>\!2$ are \textit{not} eliminated in off-axis fields endowed with differentiable AD as they are in on-axis fields endowed with non-differentiable AD. 

In our experiments, we make use of the arrangement depicted in Fig.~\ref{Fig:ConceptSetup}(f) that can introduce arbitrary AD profiles $\varphi(\omega)$, whether differentiable or non-differentiable. This pulsed-beam shaper comprises a diffraction grating and a cylindrical lens to spatially resolve and collimate the spectrum of plane-wave femtosecond pulses (Tsnumai, Spectra Physics; wavelength $\approx\!800$~nm, bandwidth $\approx\!10$~nm, and pulsewidth $\sim\!100$~fs). A reflective, phase-only SLM placed at the focal plane modulates the phase of the impinging spectrally resolved wave front. Each wavelength $\lambda$ occupies a column on the SLM, and a phase of the form $\tfrac{2\pi}{\lambda}x\sin{\varphi(\lambda)}$ deflects it at a prescribed angle $\varphi(\lambda)$ with respect to the $z$-axis. Because the SLM implements the angle $\varphi$ \textit{independently} for each $\lambda$, arbitrary functional forms for $\varphi(\lambda)$ can be produced in principle. We record $\varphi(\lambda)$ by taking a spatial Fourier transform of the spectrally resolved wave front.

We first introduce differentiable AD into on-axis fields $\varphi_{\mathrm{o}}\!=\!0$ with $\varphi_{\mathrm{o}}^{(1)}$ taking on positive and negative values [Fig.~\ref{Fig:MeasuredGVD}(a)]; i.e., $\varphi(\omega_{\mathrm{o}}+\Omega)\!=\!\varphi_{\mathrm{o}}^{(1)}\Omega$, with $\lambda_{\mathrm{o}}\!=\!799$~nm. This approximates the AD produced by a grating after eliminating higher-order AD terms $\varphi_{\mathrm{o}}^{(n)}\!=\!0$ for $n\!\geq\!2$ (see also \cite{Hall21PRA,Hall21APLP}). The GVD is anomalous whether $\varphi_{\mathrm{o}}^{(1)}$ is positive \textit{or} negative, and $\widetilde{v}\!=\!c$ at $\lambda_{\mathrm{o}}$. Because $\varphi$ is purely linear in $\Omega$, the GVD is spectrally flat. In general, this is the domain of validity of the MGF result: on-axis luminal fields endowed with differentiable AD.

We next examine the possibility of realizing \textit{normal} GVD in an on-axis field, which necessitates introducing non-differentiable AD. We show in Fig.~\ref{Fig:MeasuredGVD}(b) measurements for 5 on-axis fields whose AD are implemented according to Eq.~\ref{Eq:ADForNonDifferentiableAD} and are thus endowed with non-differentiable AD. In all cases, the field is non-luminal with $\widetilde{v}\!=\!1.5c$ and $\lambda_{\mathrm{o}}\!=\!800$~nm. The central field (shown in black) is GVD-free with $\varphi(\omega)\!\approx\!\eta\sqrt{\tfrac{\Omega}{\omega_{\mathrm{o}}}}$. Besides this field, we produce two anomalous-GVD fields $c\omega_{\mathrm{o}}k_{2}\!=\!-50$ and -100, and two normal-GVD fields with $c\omega_{\mathrm{o}}k_{2}\!=\!50$ and 100. Normal and anomalous GVD are produced here symmetrically, independently of $\widetilde{v}$. The group velocities $\widetilde{v}(\lambda)$ for all 5 fields are linear in $\lambda$ over the bandwidth used ($\Delta\lambda\!\approx\!2$~nm), except for the GVD-free field in which $\widetilde{v}(\lambda)$ is independent of $\lambda$. Consequently, $k_{z}^{(2)}$ is a wavelength-independent constant in all cases.

\begin{figure}[t!]
\centering
\includegraphics[width=8.6cm]{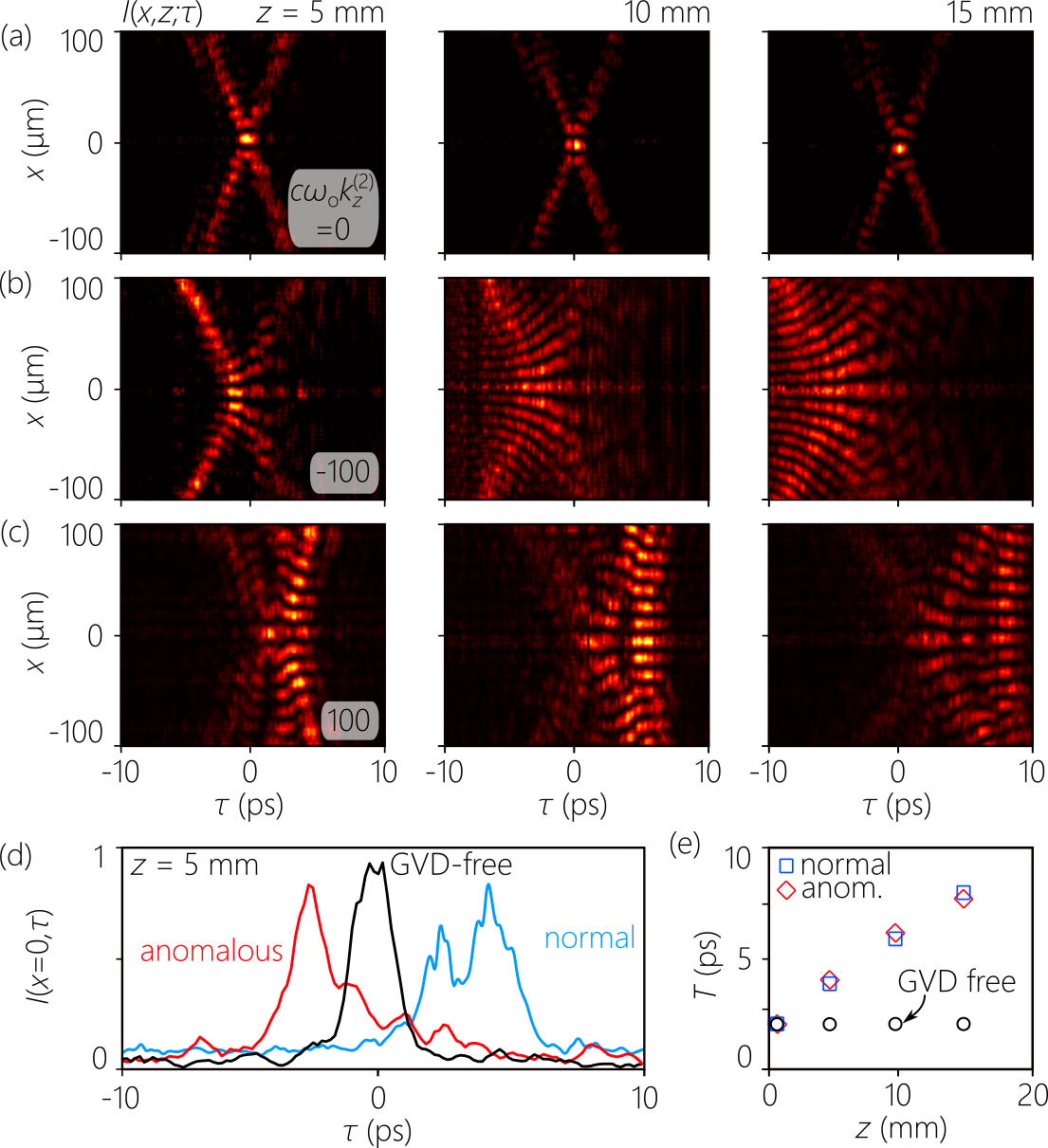}
\caption{Measured spatio-temporal profiles $I(x,z;\tau)$ all with $\widetilde{v}\!=\!1.5c$ at $z\!=\!5$, 10, and 15~mm, for (a) a propagation-invariant GVD-free field, (b) after introducing anomalous GVD $\omega_{\mathrm{o}}ck_{z}^{(2)}\!=\!-100$, and (c) normal GVD $\omega_{\mathrm{o}}ck_{z}^{(2)}\!=\!100$, all from Fig.~\ref{Fig:MeasuredGVD}(b); see Eq.~\ref{Eq:ADForNonDifferentiableAD}. At $z\!=\!0$, all three fields are identical to that in the first panel in (a). (d) The on-axis $x\!=\!0$ temporal profiles $I(0,z;\tau)$ from (a-c) at $z\!=\!5$~mm. (e) The temporal width $T$ along $z$ for the on-axis fields in (a-c).}
\label{Fig:Temporal}
\end{figure}

The third class of fields are off-axis ($\varphi_{\mathrm{o}}\!=\!1^{\circ}$) luminal configurations with $\lambda_{\mathrm{o}}\!=\!799$~nm. We control the sign of the GVD term and set $\widetilde{v}\!=\!c$ by precisely tailoring the values of $\varphi_{\mathrm{o}}$, $\varphi_{\mathrm{o}}^{(1)}$, and $\varphi_{\mathrm{o}}^{(2)}$. Setting $\delta_{\mathrm{o}}^{(1)}\!=\!-\tfrac{1}{2}\varphi_{\mathrm{o}}$ dictates that $\omega_{\mathrm{o}}\varphi_{\mathrm{o}}^{(1)}\!=\!\tan{\delta_{\mathrm{o}}^{(1)}}\!=\!-8.72\!\times\!10^{-3}$ to ensure that $\widetilde{v}\!=\!c$. The GVD coefficient is then tuned by varying the second-order AD term $\omega_{\mathrm{o}}^{2}\varphi_{\mathrm{o}}^{(2)}$. Here we use $\omega_{\mathrm{o}}^{2}\varphi_{\mathrm{o}}^{(2)}\!=\!10^{4}$ to produce anomalous GVD across the spectral range of interest, and $\omega_{\mathrm{o}}^{2}\varphi_{\mathrm{o}}^{(2)}\!=\!-10^{4}$ to produce normal GVD [Fig.~\ref{Fig:MeasuredGVD}(c)]. The contribution from the $\omega_{\mathrm{o}}^{2}\varphi_{\mathrm{o}}^{(2)}$ term dominates here over that from $\omega_{\mathrm{o}}\varphi_{\mathrm{o}}^{(1)}$, so that this scenario approaches the proposal by Porras \textit{et al}. in \cite{Porras03PRE2}, except that $\omega_{\mathrm{o}}\varphi_{\mathrm{o}}^{(1)}$ deviates slightly from 0 to adjust $\widetilde{v}$ from $c/\cos{\varphi_{\mathrm{o}}}$ to $c$.

Finally, by removing the constraint $\varphi_{\mathrm{o}}\!=\!-\tfrac{1}{2}\delta_{\mathrm{o}}^{(1)}$, the off-axis field is no longer luminal $\widetilde{v}\!\neq\!c$, and we can use the AD terms $\varphi_{\mathrm{o}}$, $\omega_{\mathrm{o}}\varphi_{\mathrm{o}}^{(1)}$ and $\omega_{\mathrm{o}}^{2}\varphi_{\mathrm{o}}^{(2)}$ to adjust the GVD [Fig.~\ref{Fig:MeasuredGVD}(d)]. We first consider a field in which $\varphi_{\mathrm{o}}\!=\!-1^{\circ}$, $\omega_{\mathrm{o}}\varphi_{\mathrm{o}}^{(1)}\!=\!-8.72$ and $\omega_{\mathrm{o}}^{2}\varphi_{\mathrm{o}}^{(2)}\!=\!10^{4}$ at $\lambda_{\mathrm{o}}\!=\!799$~nm. Because of the significant value of $\omega_{\mathrm{o}}\varphi_{\mathrm{o}}^{(1)}$ the field is no longer luminal, $\widetilde{v}\!\approx\!1.18c$, and the GVD is anomalous throughout. In another example we set $\varphi_{\mathrm{o}}\!=\!1^{\circ}$, $\omega_{\mathrm{o}}\varphi_{\mathrm{o}}^{(1)}\!=\!8.72$, and $\omega_{\mathrm{o}}^{2}\varphi_{\mathrm{o}}^{(2)}\!=\!10^{4}$. Switching the signs of both $\varphi_{\mathrm{o}}$ and $\omega_{\mathrm{o}}\varphi_{\mathrm{o}}^{(1)}$ while keeping their amplitudes constant retains $\widetilde{v}\!\approx\!1.18c$, but renders the GVD normal over part of the spectrum, as can be verified in Eq.~\ref{Eq:AxialGVDCoefficient}. Because $\varphi$ is not linear in terms of $\Omega$ in Fig.~\ref{Fig:MeasuredGVD}(c,d), the GVD is not spectrally flat.

The AD-driven GVD in free space was further verified by measuring the spatio-temporal profile of the field $I(x,z;\tau)$ along the $z$-axis by embedding the angular-dispersion synthesizer in one arm of a Mach Zehnder interferometer and placing an optical delay $\tau$ in the reference path for the original laser pulses [Fig.~\ref{Fig:ConceptSetup}(f)]. Scanning the delay $\tau$ produces spatially resolved fringes when the synthesized field overlaps with the reference pulse in space and time. We thus reconstruct $I(x,z;\tau)$ from the interference visibility in a frame moving at $\widetilde{v}\!=\!c/\widetilde{n}$ (see \cite{Kondakci19NC,Bhaduri19Optica,Yessenov19OE} for details). We plot in Fig.~\ref{Fig:Temporal} measurements for three fields from Fig.~\ref{Fig:MeasuredGVD}(b) at $z\!=\!5$, 10, and 15~mm: a propagation-invariant GVD-free field in which $\varphi(\omega)\!\propto\!\sqrt{\Omega}$ [Fig.~\ref{Fig:Temporal}(a)], and fields having anomalous [Fig.~\ref{Fig:Temporal}(b)] and normal GVD [Fig.~\ref{Fig:Temporal}(c)], which both disperse temporally, but in different directions with respect to the leading edge of the initial pulse. The on-axis pulse profiles [Fig.~\ref{Fig:Temporal}(d)] and the pulse widths along $z$ [Fig.~\ref{Fig:Temporal}(e)] both confirm the expected dispersive propagation characteristics of these two fields.

In conclusion, we have shown that the well-known result by MGF \cite{Martinez84JOSAA} implying that AD in free space yields only anomalous GVD applies in fact to only a subset of pulsed optical fields; namely, to on-axis luminal fields incorporating differentiable AD. We have demonstrated here theoretically and experimentally that normal and anomalous GVD can both be produced via AD in on-axis fields via non-differentiable AD (at non-luminal group velocities), and in off-axis fields via differentiable AD (at either luminal or non-luminal group velocities) that requires independent tunability of the first- and second-order AD terms. These results expand the repertoire of applications that can be broached by AD in the areas of dispersion compensation, pulse compression, and nonlinear optics, especially at long wavelengths where material dispersion is anomalous. Both differentiable and non-differentiable AD can be synthesized in the mid-infrared \cite{Yessenov20OSAC} by exploiting appropriate phase plates \cite{Kondakci18OE,Bhaduri19OL}. In addition, our results are a crucial step towards realizing linear propagation-invariant wave packets in the anomalous dispersion regime \cite{Longhi03PRE,Porras04PRE,Longhi04OL,Malaguti08OL,Mills12PRA}.

\section*{Funding}
U.S. Office of Naval Research (ONR) contract N00014-17-1-2458 and ONR MURI contract N00014-20-1-2789.

\vspace{2mm}
\noindent
\textbf{Disclosures.} The authors declare no conflicts of interest.

\bibliography{diffraction}

\begin{thebibliography}{38}%
\makeatletter
\providecommand \@ifxundefined [1]{%
 \@ifx{#1\undefined}
}%
\providecommand \@ifnum [1]{%
 \ifnum #1\expandafter \@firstoftwo
 \else \expandafter \@secondoftwo
 \fi
}%
\providecommand \@ifx [1]{%
 \ifx #1\expandafter \@firstoftwo
 \else \expandafter \@secondoftwo
 \fi
}%
\providecommand \natexlab [1]{#1}%
\providecommand \enquote  [1]{``#1''}%
\providecommand \bibnamefont  [1]{#1}%
\providecommand \bibfnamefont [1]{#1}%
\providecommand \citenamefont [1]{#1}%
\providecommand \href@noop [0]{\@secondoftwo}%
\providecommand \href [0]{\begingroup \@sanitize@url \@href}%
\providecommand \@href[1]{\@@startlink{#1}\@@href}%
\providecommand \@@href[1]{\endgroup#1\@@endlink}%
\providecommand \@sanitize@url [0]{\catcode `\\12\catcode `\$12\catcode
  `\&12\catcode `\#12\catcode `\^12\catcode `\_12\catcode `\%12\relax}%
\providecommand \@@startlink[1]{}%
\providecommand \@@endlink[0]{}%
\providecommand \url  [0]{\begingroup\@sanitize@url \@url }%
\providecommand \@url [1]{\endgroup\@href {#1}{\urlprefix }}%
\providecommand \urlprefix  [0]{URL }%
\providecommand \Eprint [0]{\href }%
\providecommand \doibase [0]{https://doi.org/}%
\providecommand \selectlanguage [0]{\@gobble}%
\providecommand \bibinfo  [0]{\@secondoftwo}%
\providecommand \bibfield  [0]{\@secondoftwo}%
\providecommand \translation [1]{[#1]}%
\providecommand \BibitemOpen [0]{}%
\providecommand \bibitemStop [0]{}%
\providecommand \bibitemNoStop [0]{.\EOS\space}%
\providecommand \EOS [0]{\spacefactor3000\relax}%
\providecommand \BibitemShut  [1]{\csname bibitem#1\endcsname}%
\let\auto@bib@innerbib\@empty
\bibitem [{\citenamefont {Torres}\ \emph {et~al.}(2010)\citenamefont {Torres},
  \citenamefont {Hendrych},\ and\ \citenamefont {Valencia}}]{Torres10AOP}%
  \BibitemOpen
  \bibfield  {author} {\bibinfo {author} {\bibfnamefont {J.~P.}\ \bibnamefont
  {Torres}}, \bibinfo {author} {\bibfnamefont {M.}~\bibnamefont {Hendrych}},\
  and\ \bibinfo {author} {\bibfnamefont {A.}~\bibnamefont {Valencia}},\
  }\bibfield  {title} {\bibinfo {title} {Angular dispersion: an enabling tool
  in nonlinear and quantum optics},\ }\href@noop {} {\bibfield  {journal}
  {\bibinfo  {journal} {Adv. Opt. Photon.}\ }\textbf {\bibinfo {volume} {2}},\
  \bibinfo {pages} {319} (\bibinfo {year} {2010})}\BibitemShut {NoStop}%
\bibitem [{\citenamefont {F{\"u}l{\"o}p}\ and\ \citenamefont
  {Hebling}(2010)}]{Fulop10Review}%
  \BibitemOpen
  \bibfield  {author} {\bibinfo {author} {\bibfnamefont {J.~A.}\ \bibnamefont
  {F{\"u}l{\"o}p}}\ and\ \bibinfo {author} {\bibfnamefont {J.}~\bibnamefont
  {Hebling}},\ }\bibfield  {title} {\bibinfo {title} {Applications of
  tilted-pulse-front excitation},\ }in\ \href@noop {} {\emph {\bibinfo
  {booktitle} {Recent Optical and Photonic Technologies}}},\ \bibinfo {editor}
  {edited by\ \bibinfo {editor} {\bibfnamefont {K.~Y.}\ \bibnamefont {Kim}}}\
  (\bibinfo  {publisher} {InTech},\ \bibinfo {year} {2010})\BibitemShut
  {NoStop}%
\bibitem [{\citenamefont {Fork}\ \emph {et~al.}(1984)\citenamefont {Fork},
  \citenamefont {Martinez},\ and\ \citenamefont {Gordon}}]{Fork84OL}%
  \BibitemOpen
  \bibfield  {author} {\bibinfo {author} {\bibfnamefont {R.~L.}\ \bibnamefont
  {Fork}}, \bibinfo {author} {\bibfnamefont {O.~E.}\ \bibnamefont {Martinez}},\
  and\ \bibinfo {author} {\bibfnamefont {J.~P.}\ \bibnamefont {Gordon}},\
  }\bibfield  {title} {\bibinfo {title} {Negative dispersion using pairs of
  prisms},\ }\href@noop {} {\bibfield  {journal} {\bibinfo  {journal} {Opt.
  Lett.}\ }\textbf {\bibinfo {volume} {9}},\ \bibinfo {pages} {150} (\bibinfo
  {year} {1984})}\BibitemShut {NoStop}%
\bibitem [{\citenamefont {Gordon}\ and\ \citenamefont
  {Fork}(1984)}]{Gordon84OL}%
  \BibitemOpen
  \bibfield  {author} {\bibinfo {author} {\bibfnamefont {J.~P.}\ \bibnamefont
  {Gordon}}\ and\ \bibinfo {author} {\bibfnamefont {R.~L.}\ \bibnamefont
  {Fork}},\ }\bibfield  {title} {\bibinfo {title} {Optical resonator with
  negative dispersion},\ }\href@noop {} {\bibfield  {journal} {\bibinfo
  {journal} {Opt. Lett.}\ }\textbf {\bibinfo {volume} {9}},\ \bibinfo {pages}
  {153} (\bibinfo {year} {1984})}\BibitemShut {NoStop}%
\bibitem [{\citenamefont {Lemoff}\ and\ \citenamefont
  {Barty}(1993)}]{Lemoff93OL}%
  \BibitemOpen
  \bibfield  {author} {\bibinfo {author} {\bibfnamefont {B.~E.}\ \bibnamefont
  {Lemoff}}\ and\ \bibinfo {author} {\bibfnamefont {C.~P.~J.}\ \bibnamefont
  {Barty}},\ }\bibfield  {title} {\bibinfo {title} {Quintic-phase-limited,
  spatially uniform expansion and recompression of ultrashort optical pulses},\
  }\href@noop {} {\bibfield  {journal} {\bibinfo  {journal} {Opt. Lett.}\
  }\textbf {\bibinfo {volume} {18}},\ \bibinfo {pages} {1651} (\bibinfo {year}
  {1993})}\BibitemShut {NoStop}%
\bibitem [{\citenamefont {Kane}\ and\ \citenamefont
  {Squier}(1997)}]{Kane97JOSAB}%
  \BibitemOpen
  \bibfield  {author} {\bibinfo {author} {\bibfnamefont {S.}~\bibnamefont
  {Kane}}\ and\ \bibinfo {author} {\bibfnamefont {J.}~\bibnamefont {Squier}},\
  }\bibfield  {title} {\bibinfo {title} {Grism-pair stretcher--compressor
  system for simultaneous second- and third-order dispersion compensation in
  chirped-pulse amplification},\ }\href@noop {} {\bibfield  {journal} {\bibinfo
   {journal} {J. Opt. Soc. Am. B}\ }\textbf {\bibinfo {volume} {14}},\ \bibinfo
  {pages} {661} (\bibinfo {year} {1997})}\BibitemShut {NoStop}%
\bibitem [{\citenamefont {Hebling}\ \emph {et~al.}(2002)\citenamefont
  {Hebling}, \citenamefont {Almási}, \citenamefont {Kozma},\ and\
  \citenamefont {Kuhl}}]{Hebling02OE}%
  \BibitemOpen
  \bibfield  {author} {\bibinfo {author} {\bibfnamefont {J.}~\bibnamefont
  {Hebling}}, \bibinfo {author} {\bibfnamefont {G.}~\bibnamefont {Almási}},
  \bibinfo {author} {\bibfnamefont {I.~Z.}\ \bibnamefont {Kozma}},\ and\
  \bibinfo {author} {\bibfnamefont {J.}~\bibnamefont {Kuhl}},\ }\bibfield
  {title} {\bibinfo {title} {Velocity matching by pulse front tilting for
  large-area {TH}z-pulse generation},\ }\href@noop {} {\bibfield  {journal}
  {\bibinfo  {journal} {Opt. Express}\ }\textbf {\bibinfo {volume} {10}},\
  \bibinfo {pages} {1161} (\bibinfo {year} {2002})}\BibitemShut {NoStop}%
\bibitem [{\citenamefont {Szatmari}\ \emph {et~al.}(1990)\citenamefont
  {Szatmari}, \citenamefont {Kuhnle},\ and\ \citenamefont
  {Simon}}]{Szatmari90AO}%
  \BibitemOpen
  \bibfield  {author} {\bibinfo {author} {\bibfnamefont {S.}~\bibnamefont
  {Szatmari}}, \bibinfo {author} {\bibfnamefont {G.}~\bibnamefont {Kuhnle}},\
  and\ \bibinfo {author} {\bibfnamefont {P.}~\bibnamefont {Simon}},\ }\bibfield
   {title} {\bibinfo {title} {Pulse compression and traveling wave excitation
  scheme using a single dispersive element},\ }\href@noop {} {\bibfield
  {journal} {\bibinfo  {journal} {Appl. Opt.}\ }\textbf {\bibinfo {volume}
  {29}},\ \bibinfo {pages} {5372} (\bibinfo {year} {1990})}\BibitemShut
  {NoStop}%
\bibitem [{\citenamefont {Szatm{\'a}ri}\ \emph {et~al.}(1996)\citenamefont
  {Szatm{\'a}ri}, \citenamefont {Simon},\ and\ \citenamefont
  {Feuerhake}}]{Szatmari96OL}%
  \BibitemOpen
  \bibfield  {author} {\bibinfo {author} {\bibfnamefont {S.}~\bibnamefont
  {Szatm{\'a}ri}}, \bibinfo {author} {\bibfnamefont {P.}~\bibnamefont
  {Simon}},\ and\ \bibinfo {author} {\bibfnamefont {M.}~\bibnamefont
  {Feuerhake}},\ }\bibfield  {title} {\bibinfo {title}
  {Group-velocity-dispersion-compensated propagation of short pulses in
  dispersive media},\ }\href@noop {} {\bibfield  {journal} {\bibinfo  {journal}
  {Opt. Lett.}\ }\textbf {\bibinfo {volume} {21}},\ \bibinfo {pages} {1156}
  (\bibinfo {year} {1996})}\BibitemShut {NoStop}%
\bibitem [{\citenamefont {S{\~o}najalg}\ and\ \citenamefont
  {Saari}(1996)}]{Sonajalg96OL}%
  \BibitemOpen
  \bibfield  {author} {\bibinfo {author} {\bibfnamefont {H.}~\bibnamefont
  {S{\~o}najalg}}\ and\ \bibinfo {author} {\bibfnamefont {P.}~\bibnamefont
  {Saari}},\ }\bibfield  {title} {\bibinfo {title} {Suppression of temporal
  spread of ultrashort pulses in dispersive media by bessel beam generators},\
  }\href@noop {} {\bibfield  {journal} {\bibinfo  {journal} {Opt. Lett.}\
  }\textbf {\bibinfo {volume} {21}},\ \bibinfo {pages} {1162} (\bibinfo {year}
  {1996})}\BibitemShut {NoStop}%
\bibitem [{\citenamefont {S{\~o}najalg}\ \emph {et~al.}(1997)\citenamefont
  {S{\~o}najalg}, \citenamefont {R{\"a}tsep},\ and\ \citenamefont
  {Saari}}]{Sonajalg97OL}%
  \BibitemOpen
  \bibfield  {author} {\bibinfo {author} {\bibfnamefont {H.}~\bibnamefont
  {S{\~o}najalg}}, \bibinfo {author} {\bibfnamefont {M.}~\bibnamefont
  {R{\"a}tsep}},\ and\ \bibinfo {author} {\bibfnamefont {P.}~\bibnamefont
  {Saari}},\ }\bibfield  {title} {\bibinfo {title} {Demonstration of the
  {B}essel-{X} pulse propagating with strong lateral and longitudinal
  localization in a dispersive medium},\ }\href@noop {} {\bibfield  {journal}
  {\bibinfo  {journal} {Opt. Lett.}\ }\textbf {\bibinfo {volume} {22}},\
  \bibinfo {pages} {310} (\bibinfo {year} {1997})}\BibitemShut {NoStop}%
\bibitem [{\citenamefont {Martinez}\ \emph {et~al.}(1984)\citenamefont
  {Martinez}, \citenamefont {Gordon},\ and\ \citenamefont
  {Fork}}]{Martinez84JOSAA}%
  \BibitemOpen
  \bibfield  {author} {\bibinfo {author} {\bibfnamefont {O.~E.}\ \bibnamefont
  {Martinez}}, \bibinfo {author} {\bibfnamefont {J.~P.}\ \bibnamefont
  {Gordon}},\ and\ \bibinfo {author} {\bibfnamefont {R.~L.}\ \bibnamefont
  {Fork}},\ }\bibfield  {title} {\bibinfo {title} {Negative group-velocity
  dispersion using refraction},\ }\href@noop {} {\bibfield  {journal} {\bibinfo
   {journal} {J. Opt. Soc. Am. A}\ }\textbf {\bibinfo {volume} {1}},\ \bibinfo
  {pages} {1003} (\bibinfo {year} {1984})}\BibitemShut {NoStop}%
\bibitem [{\citenamefont {Porras}\ \emph {et~al.}(2003)\citenamefont {Porras},
  \citenamefont {Valiulis},\ and\ \citenamefont {{Di T}rapani}}]{Porras03PRE2}%
  \BibitemOpen
  \bibfield  {author} {\bibinfo {author} {\bibfnamefont {M.~A.}\ \bibnamefont
  {Porras}}, \bibinfo {author} {\bibfnamefont {G.}~\bibnamefont {Valiulis}},\
  and\ \bibinfo {author} {\bibfnamefont {P.}~\bibnamefont {{Di T}rapani}},\
  }\bibfield  {title} {\bibinfo {title} {Unified description of {B}essel {X}
  waves with cone dispersion and tilted pulses},\ }\href@noop {} {\bibfield
  {journal} {\bibinfo  {journal} {Phys. Rev. E}\ }\textbf {\bibinfo {volume}
  {68}},\ \bibinfo {pages} {016613} (\bibinfo {year} {2003})}\BibitemShut
  {NoStop}%
\bibitem [{\citenamefont {Zamboni-Rached}\ \emph {et~al.}(2003)\citenamefont
  {Zamboni-Rached}, \citenamefont {N{\'o}brega}, \citenamefont
  {Hern{\'a}ndez-Figueroa},\ and\ \citenamefont {Recami}}]{Zamboni03OC}%
  \BibitemOpen
  \bibfield  {author} {\bibinfo {author} {\bibfnamefont {M.}~\bibnamefont
  {Zamboni-Rached}}, \bibinfo {author} {\bibfnamefont {K.~Z.}\ \bibnamefont
  {N{\'o}brega}}, \bibinfo {author} {\bibfnamefont {H.~E.}\ \bibnamefont
  {Hern{\'a}ndez-Figueroa}},\ and\ \bibinfo {author} {\bibfnamefont
  {E.}~\bibnamefont {Recami}},\ }\bibfield  {title} {\bibinfo {title}
  {Localized superluminal solutions to the wave equation in (vacuum or)
  dispersive media, for arbitrary frequencies and with adjustable bandwidth},\
  }\href@noop {} {\bibfield  {journal} {\bibinfo  {journal} {Opt. Commun.}\
  }\textbf {\bibinfo {volume} {226}},\ \bibinfo {pages} {15} (\bibinfo {year}
  {2003})}\BibitemShut {NoStop}%
\bibitem [{\citenamefont {Longhi}(2003)}]{Longhi03PRE}%
  \BibitemOpen
  \bibfield  {author} {\bibinfo {author} {\bibfnamefont {S.}~\bibnamefont
  {Longhi}},\ }\bibfield  {title} {\bibinfo {title} {Spatial-temporal
  {G}auss-{L}aguerre waves in dispersive media},\ }\href@noop {} {\bibfield
  {journal} {\bibinfo  {journal} {Phys. Rev. E}\ }\textbf {\bibinfo {volume}
  {68}},\ \bibinfo {pages} {066612} (\bibinfo {year} {2003})}\BibitemShut
  {NoStop}%
\bibitem [{\citenamefont {Longhi}(2004)}]{Longhi04OL}%
  \BibitemOpen
  \bibfield  {author} {\bibinfo {author} {\bibfnamefont {S.}~\bibnamefont
  {Longhi}},\ }\bibfield  {title} {\bibinfo {title} {Localized subluminal
  envelope pulses in dispersive media},\ }\href@noop {} {\bibfield  {journal}
  {\bibinfo  {journal} {Opt. Lett.}\ }\textbf {\bibinfo {volume} {29}},\
  \bibinfo {pages} {147} (\bibinfo {year} {2004})}\BibitemShut {NoStop}%
\bibitem [{\citenamefont {Porras}\ and\ \citenamefont {{Di
  T}rapani}(2004)}]{Porras04PRE}%
  \BibitemOpen
  \bibfield  {author} {\bibinfo {author} {\bibfnamefont {M.~A.}\ \bibnamefont
  {Porras}}\ and\ \bibinfo {author} {\bibfnamefont {P.}~\bibnamefont {{Di
  T}rapani}},\ }\bibfield  {title} {\bibinfo {title} {Localized and stationary
  light wave modes in dispersive media},\ }\href@noop {} {\bibfield  {journal}
  {\bibinfo  {journal} {Phys. Rev. E}\ }\textbf {\bibinfo {volume} {69}},\
  \bibinfo {pages} {066606} (\bibinfo {year} {2004})}\BibitemShut {NoStop}%
\bibitem [{\citenamefont {Malaguti}\ \emph {et~al.}(2008)\citenamefont
  {Malaguti}, \citenamefont {Bellanca},\ and\ \citenamefont
  {Trillo}}]{Malaguti08OL}%
  \BibitemOpen
  \bibfield  {author} {\bibinfo {author} {\bibfnamefont {S.}~\bibnamefont
  {Malaguti}}, \bibinfo {author} {\bibfnamefont {G.}~\bibnamefont {Bellanca}},\
  and\ \bibinfo {author} {\bibfnamefont {S.}~\bibnamefont {Trillo}},\
  }\bibfield  {title} {\bibinfo {title} {Two-dimensional envelope localized
  waves in the anomalous dispersion regime},\ }\href@noop {} {\bibfield
  {journal} {\bibinfo  {journal} {Opt. Lett.}\ }\textbf {\bibinfo {volume}
  {33}},\ \bibinfo {pages} {1117} (\bibinfo {year} {2008})}\BibitemShut
  {NoStop}%
\bibitem [{\citenamefont {Hall}\ \emph
  {et~al.}(2021{\natexlab{a}})\citenamefont {Hall}, \citenamefont {Yessenov},\
  and\ \citenamefont {Abouraddy}}]{Hall21OL}%
  \BibitemOpen
  \bibfield  {author} {\bibinfo {author} {\bibfnamefont {L.~A.}\ \bibnamefont
  {Hall}}, \bibinfo {author} {\bibfnamefont {M.}~\bibnamefont {Yessenov}},\
  and\ \bibinfo {author} {\bibfnamefont {A.~F.}\ \bibnamefont {Abouraddy}},\
  }\bibfield  {title} {\bibinfo {title} {Space-time wave packets violate the
  universal relationship between angular dispersion and pulse-front tilt},\
  }\href@noop {} {\bibfield  {journal} {\bibinfo  {journal} {Opt. Lett.}\
  }\textbf {\bibinfo {volume} {46}},\ \bibinfo {pages} {1672} (\bibinfo {year}
  {2021}{\natexlab{a}})}\BibitemShut {NoStop}%
\bibitem [{\citenamefont {Yessenov}\ \emph {et~al.}(2021)\citenamefont
  {Yessenov}, \citenamefont {Hall},\ and\ \citenamefont
  {Abouraddy}}]{Yessenov21ACSP}%
  \BibitemOpen
  \bibfield  {author} {\bibinfo {author} {\bibfnamefont {M.}~\bibnamefont
  {Yessenov}}, \bibinfo {author} {\bibfnamefont {L.~A.}\ \bibnamefont {Hall}},\
  and\ \bibinfo {author} {\bibfnamefont {A.~F.}\ \bibnamefont {Abouraddy}},\
  }\bibfield  {title} {\bibinfo {title} {Engineering the optical vacuum:
  {A}rbitrary magnitude, sign, and order of dispersion in free space using
  space-time wave packets},\ }\href@noop {} {\bibfield  {journal} {\bibinfo
  {journal} {arXiv:2102.09443}\ } (\bibinfo {year} {2021})}\BibitemShut
  {NoStop}%
\bibitem [{\citenamefont {Kondakci}\ and\ \citenamefont
  {Abouraddy}(2016)}]{Kondakci16OE}%
  \BibitemOpen
  \bibfield  {author} {\bibinfo {author} {\bibfnamefont {H.~E.}\ \bibnamefont
  {Kondakci}}\ and\ \bibinfo {author} {\bibfnamefont {A.~F.}\ \bibnamefont
  {Abouraddy}},\ }\bibfield  {title} {\bibinfo {title} {Diffraction-free pulsed
  optical beams via space-time correlations},\ }\href@noop {} {\bibfield
  {journal} {\bibinfo  {journal} {Opt. Express}\ }\textbf {\bibinfo {volume}
  {24}},\ \bibinfo {pages} {28659} (\bibinfo {year} {2016})}\BibitemShut
  {NoStop}%
\bibitem [{\citenamefont {Parker}\ and\ \citenamefont
  {Alonso}(2016)}]{Parker16OE}%
  \BibitemOpen
  \bibfield  {author} {\bibinfo {author} {\bibfnamefont {K.~J.}\ \bibnamefont
  {Parker}}\ and\ \bibinfo {author} {\bibfnamefont {M.~A.}\ \bibnamefont
  {Alonso}},\ }\bibfield  {title} {\bibinfo {title} {The longitudinal iso-phase
  condition and needle pulses},\ }\href@noop {} {\bibfield  {journal} {\bibinfo
   {journal} {Opt. Express}\ }\textbf {\bibinfo {volume} {24}},\ \bibinfo
  {pages} {28669} (\bibinfo {year} {2016})}\BibitemShut {NoStop}%
\bibitem [{\citenamefont {Kondakci}\ and\ \citenamefont
  {Abouraddy}(2017)}]{Kondakci17NP}%
  \BibitemOpen
  \bibfield  {author} {\bibinfo {author} {\bibfnamefont {H.~E.}\ \bibnamefont
  {Kondakci}}\ and\ \bibinfo {author} {\bibfnamefont {A.~F.}\ \bibnamefont
  {Abouraddy}},\ }\bibfield  {title} {\bibinfo {title} {Diffraction-free
  space-time beams},\ }\href@noop {} {\bibfield  {journal} {\bibinfo  {journal}
  {Nat. Photon.}\ }\textbf {\bibinfo {volume} {11}},\ \bibinfo {pages} {733}
  (\bibinfo {year} {2017})}\BibitemShut {NoStop}%
\bibitem [{\citenamefont {Porras}(2017)}]{Porras17OL}%
  \BibitemOpen
  \bibfield  {author} {\bibinfo {author} {\bibfnamefont {M.~A.}\ \bibnamefont
  {Porras}},\ }\bibfield  {title} {\bibinfo {title} {Gaussian beams diffracting
  in time},\ }\href@noop {} {\bibfield  {journal} {\bibinfo  {journal} {Opt.
  Lett.}\ }\textbf {\bibinfo {volume} {42}},\ \bibinfo {pages} {4679} (\bibinfo
  {year} {2017})}\BibitemShut {NoStop}%
\bibitem [{\citenamefont {Wong}\ and\ \citenamefont
  {Kaminer}(2017)}]{Wong17ACSP2}%
  \BibitemOpen
  \bibfield  {author} {\bibinfo {author} {\bibfnamefont {L.~J.}\ \bibnamefont
  {Wong}}\ and\ \bibinfo {author} {\bibfnamefont {I.}~\bibnamefont {Kaminer}},\
  }\bibfield  {title} {\bibinfo {title} {Ultrashort tilted-pulsefront pulses
  and nonparaxial tilted-phase-front beams},\ }\href@noop {} {\bibfield
  {journal} {\bibinfo  {journal} {ACS Photon.}\ }\textbf {\bibinfo {volume}
  {4}},\ \bibinfo {pages} {2257} (\bibinfo {year} {2017})}\BibitemShut
  {NoStop}%
\bibitem [{\citenamefont {Efremidis}(2017)}]{Efremidis17OL}%
  \BibitemOpen
  \bibfield  {author} {\bibinfo {author} {\bibfnamefont {N.~K.}\ \bibnamefont
  {Efremidis}},\ }\bibfield  {title} {\bibinfo {title} {Spatiotemporal
  diffraction-free pulsed beams in free-space of the {A}iry and {B}essel
  type},\ }\href@noop {} {\bibfield  {journal} {\bibinfo  {journal} {Opt.
  Lett.}\ }\textbf {\bibinfo {volume} {42}},\ \bibinfo {pages} {5038} (\bibinfo
  {year} {2017})}\BibitemShut {NoStop}%
\bibitem [{\citenamefont {Yessenov}\ \emph
  {et~al.}(2019{\natexlab{a}})\citenamefont {Yessenov}, \citenamefont
  {Bhaduri}, \citenamefont {Kondakci},\ and\ \citenamefont
  {Abouraddy}}]{Yessenov19OPN}%
  \BibitemOpen
  \bibfield  {author} {\bibinfo {author} {\bibfnamefont {M.}~\bibnamefont
  {Yessenov}}, \bibinfo {author} {\bibfnamefont {B.}~\bibnamefont {Bhaduri}},
  \bibinfo {author} {\bibfnamefont {H.~E.}\ \bibnamefont {Kondakci}},\ and\
  \bibinfo {author} {\bibfnamefont {A.~F.}\ \bibnamefont {Abouraddy}},\
  }\bibfield  {title} {\bibinfo {title} {Weaving the rainbow: Space-time
  optical wave packets},\ }\href@noop {} {\bibfield  {journal} {\bibinfo
  {journal} {Opt. Photon. News}\ }\textbf {\bibinfo {volume} {30}},\ \bibinfo
  {pages} {34} (\bibinfo {year} {2019}{\natexlab{a}})}\BibitemShut {NoStop}%
\bibitem [{\citenamefont {Yessenov}\ \emph
  {et~al.}(2019{\natexlab{b}})\citenamefont {Yessenov}, \citenamefont
  {Bhaduri}, \citenamefont {Mach}, \citenamefont {Mardani}, \citenamefont
  {Kondakci}, \citenamefont {Alonso}, \citenamefont {Atia},\ and\ \citenamefont
  {Abouraddy}}]{Yessenov19OE}%
  \BibitemOpen
  \bibfield  {author} {\bibinfo {author} {\bibfnamefont {M.}~\bibnamefont
  {Yessenov}}, \bibinfo {author} {\bibfnamefont {B.}~\bibnamefont {Bhaduri}},
  \bibinfo {author} {\bibfnamefont {L.}~\bibnamefont {Mach}}, \bibinfo {author}
  {\bibfnamefont {D.}~\bibnamefont {Mardani}}, \bibinfo {author} {\bibfnamefont
  {H.~E.}\ \bibnamefont {Kondakci}}, \bibinfo {author} {\bibfnamefont {M.~A.}\
  \bibnamefont {Alonso}}, \bibinfo {author} {\bibfnamefont {G.~A.}\
  \bibnamefont {Atia}},\ and\ \bibinfo {author} {\bibfnamefont {A.~F.}\
  \bibnamefont {Abouraddy}},\ }\bibfield  {title} {\bibinfo {title} {What is
  the maximum differential group delay achievable by a space-time wave packet
  in free space?},\ }\href@noop {} {\bibfield  {journal} {\bibinfo  {journal}
  {Opt. Express}\ }\textbf {\bibinfo {volume} {27}},\ \bibinfo {pages} {12443}
  (\bibinfo {year} {2019}{\natexlab{b}})}\BibitemShut {NoStop}%
\bibitem [{\citenamefont {Yessenov}\ \emph
  {et~al.}(2019{\natexlab{c}})\citenamefont {Yessenov}, \citenamefont
  {Bhaduri}, \citenamefont {Kondakci},\ and\ \citenamefont
  {Abouraddy}}]{Yessenov19PRA}%
  \BibitemOpen
  \bibfield  {author} {\bibinfo {author} {\bibfnamefont {M.}~\bibnamefont
  {Yessenov}}, \bibinfo {author} {\bibfnamefont {B.}~\bibnamefont {Bhaduri}},
  \bibinfo {author} {\bibfnamefont {H.~E.}\ \bibnamefont {Kondakci}},\ and\
  \bibinfo {author} {\bibfnamefont {A.~F.}\ \bibnamefont {Abouraddy}},\
  }\bibfield  {title} {\bibinfo {title} {Classification of
  propagation-invariant space-time light-sheets in free space: Theory and
  experiments},\ }\href@noop {} {\bibfield  {journal} {\bibinfo  {journal}
  {Phys. Rev. A}\ }\textbf {\bibinfo {volume} {99}},\ \bibinfo {pages} {023856}
  (\bibinfo {year} {2019}{\natexlab{c}})}\BibitemShut {NoStop}%
\bibitem [{\citenamefont {Hebling}(1996)}]{Hebling96OQE}%
  \BibitemOpen
  \bibfield  {author} {\bibinfo {author} {\bibfnamefont {J.}~\bibnamefont
  {Hebling}},\ }\bibfield  {title} {\bibinfo {title} {Derivation of the pulse
  front tilt caused by angular dispersion},\ }\href@noop {} {\bibfield
  {journal} {\bibinfo  {journal} {Opt. Quant. Electron.}\ }\textbf {\bibinfo
  {volume} {28}},\ \bibinfo {pages} {1759} (\bibinfo {year}
  {1996})}\BibitemShut {NoStop}%
\bibitem [{\citenamefont {Hall}\ and\ \citenamefont
  {Abouraddy}(2021)}]{Hall21PRA}%
  \BibitemOpen
  \bibfield  {author} {\bibinfo {author} {\bibfnamefont {L.~A.}\ \bibnamefont
  {Hall}}\ and\ \bibinfo {author} {\bibfnamefont {A.~F.}\ \bibnamefont
  {Abouraddy}},\ }\bibfield  {title} {\bibinfo {title} {Free-space
  group-velocity dispersion induced in space-time wave packets by {V}-shaped
  spectra},\ }\href@noop {} {\bibfield  {journal} {\bibinfo  {journal} {Phys.
  Rev. A}\ }\textbf {\bibinfo {volume} {104}},\ \bibinfo {pages} {013505}
  (\bibinfo {year} {2021})}\BibitemShut {NoStop}%
\bibitem [{\citenamefont {Hall}\ \emph
  {et~al.}(2021{\natexlab{b}})\citenamefont {Hall}, \citenamefont {Yessenov},
  \citenamefont {Ponomarenko},\ and\ \citenamefont {Abouraddy}}]{Hall21APLP}%
  \BibitemOpen
  \bibfield  {author} {\bibinfo {author} {\bibfnamefont {L.~A.}\ \bibnamefont
  {Hall}}, \bibinfo {author} {\bibfnamefont {M.}~\bibnamefont {Yessenov}},
  \bibinfo {author} {\bibfnamefont {S.~A.}\ \bibnamefont {Ponomarenko}},\ and\
  \bibinfo {author} {\bibfnamefont {A.~F.}\ \bibnamefont {Abouraddy}},\
  }\bibfield  {title} {\bibinfo {title} {The space-time {T}albot effect},\
  }\href@noop {} {\bibfield  {journal} {\bibinfo  {journal} {APL Photon.}\
  }\textbf {\bibinfo {volume} {6}},\ \bibinfo {pages} {056105} (\bibinfo {year}
  {2021}{\natexlab{b}})}\BibitemShut {NoStop}%
\bibitem [{\citenamefont {Kondakci}\ and\ \citenamefont
  {Abouraddy}(2019)}]{Kondakci19NC}%
  \BibitemOpen
  \bibfield  {author} {\bibinfo {author} {\bibfnamefont {H.~E.}\ \bibnamefont
  {Kondakci}}\ and\ \bibinfo {author} {\bibfnamefont {A.~F.}\ \bibnamefont
  {Abouraddy}},\ }\bibfield  {title} {\bibinfo {title} {Optical space-time wave
  packets of arbitrary group velocity in free space},\ }\href@noop {}
  {\bibfield  {journal} {\bibinfo  {journal} {Nat. Commun.}\ }\textbf {\bibinfo
  {volume} {10}},\ \bibinfo {pages} {929} (\bibinfo {year} {2019})}\BibitemShut
  {NoStop}%
\bibitem [{\citenamefont {Bhaduri}\ \emph
  {et~al.}(2019{\natexlab{a}})\citenamefont {Bhaduri}, \citenamefont
  {Yessenov},\ and\ \citenamefont {Abouraddy}}]{Bhaduri19Optica}%
  \BibitemOpen
  \bibfield  {author} {\bibinfo {author} {\bibfnamefont {B.}~\bibnamefont
  {Bhaduri}}, \bibinfo {author} {\bibfnamefont {M.}~\bibnamefont {Yessenov}},\
  and\ \bibinfo {author} {\bibfnamefont {A.~F.}\ \bibnamefont {Abouraddy}},\
  }\bibfield  {title} {\bibinfo {title} {Space-time wave packets that travel in
  optical materials at the speed of light in vacuum},\ }\href@noop {}
  {\bibfield  {journal} {\bibinfo  {journal} {Optica}\ }\textbf {\bibinfo
  {volume} {6}},\ \bibinfo {pages} {139} (\bibinfo {year}
  {2019}{\natexlab{a}})}\BibitemShut {NoStop}%
\bibitem [{\citenamefont {Yessenov}\ \emph {et~al.}(2020)\citenamefont
  {Yessenov}, \citenamefont {Ru}, \citenamefont {Schepler}, \citenamefont
  {Meem}, \citenamefont {Menon}, \citenamefont {Vodopyanov},\ and\
  \citenamefont {Abouraddy}}]{Yessenov20OSAC}%
  \BibitemOpen
  \bibfield  {author} {\bibinfo {author} {\bibfnamefont {M.}~\bibnamefont
  {Yessenov}}, \bibinfo {author} {\bibfnamefont {Q.}~\bibnamefont {Ru}},
  \bibinfo {author} {\bibfnamefont {K.~L.}\ \bibnamefont {Schepler}}, \bibinfo
  {author} {\bibfnamefont {M.}~\bibnamefont {Meem}}, \bibinfo {author}
  {\bibfnamefont {R.}~\bibnamefont {Menon}}, \bibinfo {author} {\bibfnamefont
  {K.~L.}\ \bibnamefont {Vodopyanov}},\ and\ \bibinfo {author} {\bibfnamefont
  {A.~F.}\ \bibnamefont {Abouraddy}},\ }\bibfield  {title} {\bibinfo {title}
  {Mid-infrared diffraction-free space-time wave packets},\ }\href@noop {}
  {\bibfield  {journal} {\bibinfo  {journal} {OSA Continuum}\ }\textbf
  {\bibinfo {volume} {3}},\ \bibinfo {pages} {420} (\bibinfo {year}
  {2020})}\BibitemShut {NoStop}%
\bibitem [{\citenamefont {Kondakci}\ \emph {et~al.}(2018)\citenamefont
  {Kondakci}, \citenamefont {Yessenov}, \citenamefont {Meem}, \citenamefont
  {Reyes}, \citenamefont {Thul}, \citenamefont {Fairchild}, \citenamefont
  {Richardson}, \citenamefont {Menon},\ and\ \citenamefont
  {Abouraddy}}]{Kondakci18OE}%
  \BibitemOpen
  \bibfield  {author} {\bibinfo {author} {\bibfnamefont {H.~E.}\ \bibnamefont
  {Kondakci}}, \bibinfo {author} {\bibfnamefont {M.}~\bibnamefont {Yessenov}},
  \bibinfo {author} {\bibfnamefont {M.}~\bibnamefont {Meem}}, \bibinfo {author}
  {\bibfnamefont {D.}~\bibnamefont {Reyes}}, \bibinfo {author} {\bibfnamefont
  {D.}~\bibnamefont {Thul}}, \bibinfo {author} {\bibfnamefont {S.~R.}\
  \bibnamefont {Fairchild}}, \bibinfo {author} {\bibfnamefont {M.}~\bibnamefont
  {Richardson}}, \bibinfo {author} {\bibfnamefont {R.}~\bibnamefont {Menon}},\
  and\ \bibinfo {author} {\bibfnamefont {A.~F.}\ \bibnamefont {Abouraddy}},\
  }\bibfield  {title} {\bibinfo {title} {Synthesizing broadband
  propagation-invariant space-time wave packets using transmissive phase
  plates},\ }\href@noop {} {\bibfield  {journal} {\bibinfo  {journal} {Opt.
  Express}\ }\textbf {\bibinfo {volume} {26}},\ \bibinfo {pages} {13628}
  (\bibinfo {year} {2018})}\BibitemShut {NoStop}%
\bibitem [{\citenamefont {Bhaduri}\ \emph
  {et~al.}(2019{\natexlab{b}})\citenamefont {Bhaduri}, \citenamefont
  {Yessenov}, \citenamefont {Reyes}, \citenamefont {Pena}, \citenamefont
  {Meem}, \citenamefont {Fairchild}, \citenamefont {Menon}, \citenamefont
  {Richardson},\ and\ \citenamefont {Abouraddy}}]{Bhaduri19OL}%
  \BibitemOpen
  \bibfield  {author} {\bibinfo {author} {\bibfnamefont {B.}~\bibnamefont
  {Bhaduri}}, \bibinfo {author} {\bibfnamefont {M.}~\bibnamefont {Yessenov}},
  \bibinfo {author} {\bibfnamefont {D.}~\bibnamefont {Reyes}}, \bibinfo
  {author} {\bibfnamefont {J.}~\bibnamefont {Pena}}, \bibinfo {author}
  {\bibfnamefont {M.}~\bibnamefont {Meem}}, \bibinfo {author} {\bibfnamefont
  {S.~R.}\ \bibnamefont {Fairchild}}, \bibinfo {author} {\bibfnamefont
  {R.}~\bibnamefont {Menon}}, \bibinfo {author} {\bibfnamefont {M.~C.}\
  \bibnamefont {Richardson}},\ and\ \bibinfo {author} {\bibfnamefont {A.~F.}\
  \bibnamefont {Abouraddy}},\ }\bibfield  {title} {\bibinfo {title} {Broadband
  space-time wave packets propagating 70~m},\ }\href@noop {} {\bibfield
  {journal} {\bibinfo  {journal} {Opt. Lett.}\ }\textbf {\bibinfo {volume}
  {44}},\ \bibinfo {pages} {2073} (\bibinfo {year}
  {2019}{\natexlab{b}})}\BibitemShut {NoStop}%
\bibitem [{\citenamefont {Mills}\ \emph {et~al.}(2012)\citenamefont {Mills},
  \citenamefont {Siviloglou}, \citenamefont {Efremidis}, \citenamefont {Graf},
  \citenamefont {Wright}, \citenamefont {Moloney},\ and\ \citenamefont
  {Christodoulides}}]{Mills12PRA}%
  \BibitemOpen
  \bibfield  {author} {\bibinfo {author} {\bibfnamefont {M.~S.}\ \bibnamefont
  {Mills}}, \bibinfo {author} {\bibfnamefont {G.~A.}\ \bibnamefont
  {Siviloglou}}, \bibinfo {author} {\bibfnamefont {N.}~\bibnamefont
  {Efremidis}}, \bibinfo {author} {\bibfnamefont {T.}~\bibnamefont {Graf}},
  \bibinfo {author} {\bibfnamefont {E.~M.}\ \bibnamefont {Wright}}, \bibinfo
  {author} {\bibfnamefont {J.~V.}\ \bibnamefont {Moloney}},\ and\ \bibinfo
  {author} {\bibfnamefont {D.~N.}\ \bibnamefont {Christodoulides}},\ }\bibfield
   {title} {\bibinfo {title} {Localized waves with spherical harmonic
  symmetries},\ }\href@noop {} {\bibfield  {journal} {\bibinfo  {journal}
  {Phys. Rev. A}\ }\textbf {\bibinfo {volume} {86}},\ \bibinfo {pages} {063811}
  (\bibinfo {year} {2012})}\BibitemShut {NoStop}%
\end{thebibliography}%

\end{document}